\documentclass[12pt,reqno]{article}
\usepackage{amsfonts}
\usepackage{amsmath}
\usepackage{amsbsy}

\usepackage{amssymb,latexsym}
\numberwithin{equation}{section}

\begin{document}
 \allowdisplaybreaks[1]
\title{Dualisation of the D=7 Heterotic String}
\author{Nejat T. Y$\i$lmaz\\
          Department of Physics,\\
          Middle East Technical University,\\
          06531 Ankara, Turkey.\\
          \texttt{ntyilmaz@metu.edu.tr}}
\maketitle
\begin{abstract}The dualisation and the first-order formulation of the
$D=7$ abelian Yang-Mills supergravity which is the low energy
effective limit of the $D=7$ fully Higssed heterotic string is
discussed. The non-linear coset formulation of the scalars is
enlarged to include the entire bosonic sector by introducing dual
fields and by constructing the Lie superalgebra which generates
the dualized coset element.

\end{abstract}

\section{Introduction}
The supergravity theory which has the highest spacetime dimension
is the $D=11$, $N=1$ supergravity \cite{d=11}. There are three
types of supergravity theories in ten-dimensions namely the IIA,
\cite{2A1,2A2,2A3}, the IIB, \cite{2B1,2B2,2B3} and as the third
supergravity, the ten-dimensional type I supergravity theory which
is coupled to the Yang-Mills theory \cite{d=10,tani15}. One can
obtain the $D=10$, IIA supergravity theory by the Kaluza-Klein
dimensional reduction of the $D=11$ supergravity on the circle,
$S^{1}$. The supergravity theories for $D<10$ dimensions can be
obtained from the $D=11$ and the $D=10$ supergravities by the
dimensional reduction and the truncation of fields. The ten
dimensional IIA and the IIB supergravity theories are the massless
sectors or the low energy effective limits of the type IIA and the
type IIB superstring theories respectively. The type I Yang-Mills
supergravity theory in ten-dimensions on the other hand is the low
energy effective limit of the type I superstring theory and the
heterotic string theory. The eleven dimensional supergravity is
conjectured to be the low energy effective theory of the eleven
dimensional M theory.

The global symmetries of the supergravities which are obtained
from the $D=11$ supergravity \cite{d=11} by torodial
compactification and by partial dualisation of the fields, also
the coset formulation of the scalar sectors of these theories are
studied in \cite{julia1}. The complete dualisation and thus the
non-linear realization of the bosonic sectors of the IIB and the
maximal supergravities which can be obtained by the Kaluza-Klein
dimensional reduction of the $D=11$ supergravity \cite{d=11} over
the tori $T^{n}$ are given in \cite{julia2}. These supergravities
have scalar coset manifolds $G/K$ which are based on split real
form global symmetry groups $G$ \cite{hel,ker1,nej1}. The
dualisation of a generic scalar coset which has a split real form
global symmetry group is studied in \cite{nej1} whereas this
formulation is generalized to the non-split scalar cosets in
\cite{nej2}. Based on the analysis of \cite{nej2}, the non-linear
realization of the $D=8$ Salam-Sezgin matter coupled supergravity
which has a non-split scalar coset is studied in \cite{nej3}. A
general dualisation treatment of the matter coupled non-split
scalar cosets is also given \cite{nej4}.

The low energy effective limit of the seven dimensional heterotic
string theory is the $D=7$ supergravity \cite{town1,town2} which
is coupled to $19$ vector supermultiplets \cite{town3}. The $D=7$
matter coupled supergravity is constructed in \cite{koh} for an
arbitrary number of vector supermultiplets. In this work we
perform the complete bosonic dualisation of the $D=7$ matter
coupled supergravity which is an Abelian Yang-Mills supergravity
\cite{koh} and which corresponds to the low energy effective limit
of the fully Higgsed $D=7$ heterotic string. Likewise in
\cite{julia2} the dualisation of the bosonic fields will lead to
the construction of the bosonic sector of the $D=7$ Abelian
Yang-Mills supergravity as a non-linear coset model and we will
also obtain the first-order formulation of the theory.

In section two we will discuss the coset formulation of the scalar
sector of the $D=7$ matter coupled supergravity \cite{koh} also
after deriving the bosonic field equations we will locally
integrate them to obtain the first-order field equations. In
section three by following the dualisation method of \cite{julia2}
we will introduce dual fields (Lagrange multipliers \cite{pope})
for the bosonic field content and construct the Lie superalgebra
which generates the dualized coset element whose Cartan form
realizes the original second-order field equations by satisfying
the Cartan-Maurer equation and the first-order field equations by
satisfying a twisted self-duality condition
\cite{julia2,nej1,nej2,nej3,nej4}. Since the scalar coset of the
$D=7$ Abelian Yang-Mills supergravity is in general a non-split
type depending on $N$ (the number of the coupling vector
multiplets) we will effectively use the results of \cite{nej2}
which studies the dualisation of the non-split scalar cosets.
\section{The Effective D=7 Heterotic String}
 The matter coupled
$D=7$ supergravity can be obtained \cite{tez21} by the
$T^{3}$-compactification of the $D=10$ type I supergravity that is
coupled to the Yang-Mills theory \cite{d=10,tani15}. The ten
dimensional type I supergravity which is coupled to $16$ vector
supermultiplets is the low energy effective limit of the $D=10$
heterotic string. Thus the $D=7$ matter coupled Yang-Mills
supergravity \cite{koh} is the low energy effective limit of the
$D=7$ heterotic string theory \cite{town3}. We will consider the
fully Higgsed vacuum configuration for the heterotic string which
causes a spontaneous symmetry breakdown of the full Yang-Mills
gauge symmetry to its maximal torus subgroup. The low energy
effective limit of the fully Higgsed $D=10$ heterotic string is
the $U(1)^{16}$ Abelian Yang-Mills supergravity where the
$E_{8}\times E_{8}$ gauge symmetry is replaced by its maximal
torus subgroup $U(1)^{16}$ whose Lie algebra is the Cartan
subalgebra of $E_{8}\times E_{8}$. For the lower dimensional
theories one may consider the dimensional reduction of the $D=10$
Abelian Yang-Mills supergravity which is the maximal torus
subtheory of the $D=10$ Yang-Mills supergravity \cite{tez21}.
Therefore the $D=7$ Abelian Yang-Mills supergravity \cite{koh}
corresponds to the low energy effective limit of the fully Higssed
$D=7$ heterotic string. We will follow a general formulation and
consider an arbitrary number $N$ of coupling vector multiplets
bearing in mind that if one chooses $N=19$ one obtains the
effective fully Higssed $D=7$ heterotic string.

The original field content of the seven dimensional pure
$\mathcal{N}=2$ supergravity consists of a siebenbein
$e_{\mu}^{m}$, $Sp(1)$ pseudo-Majorana gravitinos $\psi_{\mu}^{i}$
for $i=1,2$, an $Sp(1)$ pseudo-Majorana spinor $\chi^{k}$, for
$j=1,2,3$ a triplet of one-forms $\{A^{j}_{\mu}\}$, a two-form
field $B_{\mu\nu}$ and a dilaton $\sigma$. In \cite{koh} the $D=7$
Abelian Yang-Mills supergravity is constructed by coupling to the
original fields an arbitrary number of $N$ vector multiplets
\begin{equation}\label{21}
(A_{\mu},\lambda^{i},\varphi^{\alpha}),
\end{equation}
where $A_{\mu}$ is a one-form field, $\lambda^{i}$ is an $Sp(1)$
pseudo-Majorana spinor and for $\alpha=1,2,3$ the scalars
$\varphi^{\alpha}$ form an $Sp(1)$ triplet. Thus the total field
content of the seven dimensional Abelian Yang-Mills supergravity
is
\begin{equation}\label{22}
(e_{\mu}^{m},\psi_{\mu}^{i},\chi^{k},B_{\mu\nu},\sigma,
A_{\mu}^{I},\lambda^{ai},\varphi^{\beta}),
\end{equation}
where $I=1,...,N+3$, $a=1,...,N$ and $\beta=1,...,3N$. The $3N$
scalars $\varphi^{\beta}$ of the $N$ vector multiplets
parameterize the $SO(N,3)/SO(N)\times SO(3)$ coset manifold.
$SO(N,3)$ is in general a non-compact real form of a semi-simple
Lie group and $SO(N)\times SO(3)$ is its maximal compact subgroup.
For this reason $SO(N,3)/SO(N)\times SO(3)$ is a Riemannian
globally symmetric space for all the $SO(N,3)$-invariant
Riemannian structures on $SO(N,3)/SO(N)\times SO(3)$ \cite{hel}.
Depending on the number of coupling vector multiplets $SO(N,3)$
can be a maximally non-compact (split) real form or not. We will
consider the general case of non-split real forms for $SO(N,3)$
which contains the split real forms as special cases. Therefore
the part of the bosonic Lagrangian which governs the vector
multiplet scalars $\varphi^{\beta}$ can be expressed as a general
symmetric space sigma model Lagrangian \cite{nej2}. One can use
the solvable Lie algebra parametrization \cite{fre} to
parameterize the scalar coset manifold $SO(N,3)/SO(N)\times
SO(3)$. The solvable Lie algebra parametrization is based on the
Iwasawa decomposition \cite{hel}
\begin{subequations}\label{23}
\begin{align}
so(N,3)&=\mathbf{k}_{0}\oplus \mathbf{s}_{0}\notag\\
\notag\\
&=\mathbf{k}_{0}\oplus \mathbf{h_{k}}\oplus
\mathbf{n_{k}},\tag{\ref{23}}
\end{align}
\end{subequations}
where $\mathbf{k}_{0}$ is the Lie algebra of $SO(N)\times SO(3)$
and $\mathbf{s}_{0}$ is the solvable Lie subalgebra of $so(N,3)$.
In \eqref{23} $\mathbf{h_{k}}$ is a subalgebra of the Cartan
subalgebra $\mathbf{h}_{0}$ of $so(N,3)$ and it generates the
maximal R-split torus in $SO(N,3)$ \cite{hel,nej2,fre,ker2}. The
nilpotent Lie subalgebra $\mathbf{n_{k}}$ of $so(N,3)$ is
generated by a subset $\{E_{m}\}$ of the positive root generators
of $so(N,3)$ such that $m\in\Delta_{nc}^{+}$. The roots in
$\Delta_{nc}^{+}$ are the non-compact roots with respect to the
Cartan involution $\theta$  which is induced by the Cartan
decomposition \cite{hel,nej2,nej4}
\begin{equation}\label{24}
so(N,3)=\mathbf{k}_{0}\oplus\mathbf{u}_{0}.
\end{equation}
The difeomorphism from $\mathbf{u}_{0}\simeq \mathbf{s}_{0}$ onto
the Riemannian globally symmetric space  $SO(N,3)/SO(N)\times
SO(3)$ \cite{hel} enables the construction of the solvable Lie
algebra parametrization for $SO(N,3)/SO(N)\times SO(3)$ such that
the coset representatives can be expressed as
\begin{equation}\label{25}
\begin{aligned}
L&=\mathbf{g}_{H}\mathbf{g}_{N}\\
\\
&=e^{\frac{1}{2}\phi ^{i}(x)H_{i}}e^{\chi ^{m}(x)E_{m}},
\end{aligned}
\end{equation}
where $\{H_{i}\}$ for $i=1,...,$ dim$\mathbf{h}_{k}$ are the
generators of $\mathbf{h}_{k}$ and $\{E_{m}\}$ for
$m\in\Delta_{nc}^{+}$ are the positive root generators which
generate the orthogonal complement of $\mathbf{h}_{k}$ within the
solvable Lie algebra $\mathbf{s}_{0}$ of $so(N,3)$ namely
$\mathbf{n_{k}}$. We have classified the $3N$ vector multiplet
scalars $\varphi^{\beta}$ as $\{\phi^{i}\}$, the dilatons for
$i=1,...,$ dim$\mathbf{h}_{k}$ and $\{\chi^{m}\}$, the axions for
$m\in\Delta_{nc}^{+}$. We will label the roots which are elements
of $\Delta_{nc}^{+}$ from $1$ to $n$ where $n$ is the dimension of
$\mathbf{n}_{k}$. Also we will refer to the dimension of
$\mathbf{h}_{k}$ as $r$. Thus we have $r+n=3N$. Since \eqref{25}
is a map from the seven dimensional spacetime into $SO(N,3)$
(whose range gives the representatives of the left cosets
$SO(N,3)/SO(N)\times SO(3)$ in one to one correspondence) it
satisfies the defining relations of $SO(N,3)$;
\begin{equation}\label{26}
L^{T}\eta L=\eta\quad\quad,\quad\quad L^{-1}=\eta L^{T}\eta,
\end{equation}
where  $\eta=(-,-,-,+,+,+,...)$. The coset representatives
\eqref{25} can be chosen as symmetric matrices in the
$(N+3)$-dimensional representation which is evident from the
explicit construction of the coset space $SO(N,3)/SO(N)\times
SO(3)$ \cite{tez21}. Thus we will take the coset representatives
as symmetric matrices; $L^{T}=L$ however likewise in \cite{nej3}
we will keep on using $L^{T}$ in our formulation for the sake of
accuracy. If we assume a matrix representation for the algebra
$so(N,3)$ then from \eqref{25} we have
\begin{subequations}\label{27}
\begin{gather}
\partial_{i}L\equiv\frac{\partial L}{\partial\phi^{i}}=\frac{1}{2}H_{i}L\quad\quad,\quad\quad
\partial_{i}L^{T}\equiv\frac{\partial L^{T}}{\partial\phi^{i}}=\frac{1}{2}L^{T}H_{i}^{T},\notag\\
\notag\\
\partial_{i}L^{-1}\equiv\frac{\partial L^{-1}}{\partial\phi^{i}}=-\frac{1}{2}L^{-1}H_{i}.\tag{\ref{27}}
\end{gather}
\end{subequations}
By differentiating the identities \eqref{26} with respect to the
dilatons $\phi^{i}$ also by using the identities \eqref{27}
bearing in mind that the generators $\{H_{i}\}$ are Cartan
generators and we can choose the coset representatives \eqref{25}
as symmetric matrices one can show that
\begin{equation}\label{28}
(H_{i}L)^{T}=H_{i}L\quad\quad,\quad\quad H_{i}^{T}\eta=-\eta
H_{i}.
\end{equation}
If we introduce the internal metric
\begin{equation}\label{29}
\mathcal{M}=L^{T}L,
\end{equation}
then the Lagrangian which corresponds to the $3N$ vector multiplet
scalars of the $D=7$ Abelian Yang-Mills supergravity can be given
as \cite{julia2,nej2,nej4,ker2}
\begin{equation}\label{29.5}
 \mathcal{L}_{scalar}=-\frac{1}{8}tr(\ast d\mathcal{M}^{-1}\wedge
 d\mathcal{M}).
\end{equation}
Thus now we can express the bosonic Lagrangian of the $D=7$
Abelian Yang-Mills supergravity as \cite{koh}
\begin{equation}\label{210}
\begin{aligned}
\mathcal{L}&=\frac{1}{2}R\ast1-\frac{5}{8} \ast d\sigma\wedge
d\sigma-\frac{1}{2}e^{2\sigma}\ast G
\wedge G\\
\\
&\quad -\frac{1}{8}tr( \ast d\mathcal{M}^{-1}\wedge
d\mathcal{M})-\frac{1}{2}e^{\sigma} F\wedge\mathcal{M} \ast F,
\end{aligned}
\end{equation}
where the coupling between the field strengths $F^{I}=dA^{I}$ for
$I=1,...,N+3$ and the scalars which parameterize the coset
$SO(N,3)/SO(N)\times SO(3)$ can be explicitly written as
\begin{equation}\label{211}
-\frac{1}{2}e^{\sigma} F\wedge\mathcal{M}\ast
F=-\frac{1}{2}e^{\sigma}\mathcal{M}_{ij} F^{i}\wedge \ast F^{j}.
\end{equation}
We have assumed the ($N+3$)-dimensional matrix representation of
$so(N,3)$. The Chern-Simon form $G$ is defined as \cite{koh}
\begin{equation}\label{212}
 G=dB-\frac{1}{\sqrt{2}}\eta_{ij}A^{i}\wedge F^{j}.
\end{equation}
The Lagrangian \eqref{210} can be identified as the
$T^{3}$-compactification of the $D=10$ Abelian Yang-Mills
supergravity \cite{town3,tez21} and it has an equivalent dual form
in which the two-form potential $B$ is replaced by a Lagrange
multiplier three-form field \cite{town1,town2}. This dual
Lagrangian corresponds to the $K_{3}$ compactification of the
$D=11$ supergravity \cite{townref25}.

If we vary the Lagrangian \eqref{210} with respect to the fields
$\sigma,B$ and $\{A^{i}\}$ we can find the corresponding field
equations as
\begin{subequations}\label{213}
\begin{gather}
-\frac{5}{4}d(\ast d\sigma)=-e^{2\sigma} \ast G\wedge
G-\frac{1}{2}e^{\sigma}\mathcal{M}_{ij} \ast F^{i}\wedge
F^{j},\notag\\
\notag\\
d(e^{2\sigma}\ast G)=0,\notag\\
\notag\\
d(e^{\sigma}\mathcal{M}_{ij}\ast
F^{j})=\sqrt{2}\,e^{2\sigma}\eta_{ij}F^{j}\wedge\ast
G.\tag{\ref{213}}
\end{gather}
\end{subequations}
The last equation can also be expressed as
\begin{equation}\label{214}
\begin{aligned}
d(e^{\frac{1}{2}\sigma}L\ast F)&=-\frac{1}{2}d\sigma\wedge
e^{\frac{1}{2}\sigma}L\ast F-\mathcal{G}_{0}^{T}\wedge
e^{\frac{1}{2}\sigma}L\ast F\\
\\
&\quad+\sqrt{2}\,(L^{T})^{-1}\eta \,e^{\frac{1}{2}\sigma}F\wedge
e^{\sigma}\ast G,
\end{aligned}
\end{equation}
where we have used the Cartan-Maurer form
\begin{equation}\label{215}
\mathcal{G}_{0}=dLL^{-1},
\end{equation}
which is calculated in \cite{nej2} as
\begin{equation}\label{216}
\begin{aligned}
\mathcal{G}_{0}&=\frac{1}{2}d\phi ^{i}H_{i}+e^{%
\frac{1}{2}\alpha _{i}\phi ^{i}}U^{\alpha }E_{\alpha }\\
\\
&=\frac{1}{2}d\phi ^{i}H_{i}+\overset{\rightharpoonup }{
\mathbf{E}^{\prime }}\:\mathbf{\Omega }\:\overset{\rightharpoonup
}{d\chi },
\end{aligned}
\end{equation}
where $\{H_{i}\}$ for $i=1,...,r$ are the generators of
$\mathbf{h}_{k}$ and $\{E_{\alpha}\}$ for
$\alpha\in\Delta_{nc}^{+}$ are the generators of $\mathbf{n}_{k}$
as defined before. We have introduced the column vector
\begin{equation}\label{216.5}
U^{\alpha}=\mathbf{\Omega}^{\alpha}_{\beta}d\chi^{\beta},
\end{equation}
and the row vector
\begin{equation}\label{216.6}
(\overset{\rightharpoonup }{
\mathbf{E}^{\prime }})_{\alpha}=e^{%
\frac{1}{2}\alpha _{i}\phi ^{i}}E_{\alpha},
\end{equation}
where the components of the roots $\alpha\in\Delta_{nc}^{+}$ are
defined as $[H_{i},E_{\alpha}]=\alpha_{i}E_{\alpha}$. Also
$\mathbf{\Omega}$ is an $n\times n$ matrix
\begin{equation}\label{217}
\begin{aligned}
 \mathbf{\Omega}&=\sum\limits_{m=0}^{\infty }\dfrac{\omega
^{m}}{(m+1)!}\\
\\
&=(e^{\omega}-I)\,\omega^{-1}.
\end{aligned}
\end{equation}
The matrix $\omega$ is
$\omega_{\beta}^{\gamma}=\chi^{\alpha}K_{\alpha\beta}^{\gamma}$.
The structure constants $K_{\alpha\beta}^{\gamma}$ are defined as
$[E_{\alpha},E_{\beta}]=K_{\alpha\beta}^{\gamma}E_{\gamma}$. In
other words since $[E_
{\alpha},E_{\beta}]=N_{\alpha,\beta}E_{\alpha+\beta}$ we have
$K_{\beta\beta}^{\alpha}=0$,
$K_{\beta\gamma}^{\alpha}=N_{\beta,\gamma}$ if
$\beta+\gamma=\alpha$ and $K_{\beta\gamma}^{\alpha}=0$ if
$\beta+\gamma\neq\alpha$ in the root sense.

 By following the analysis of \cite{ker1,ker2} the field equations for
$\{\phi^{i}\}$ and $\{\chi^{m}\}$ can be found as
\begin{equation}\label{218}
\begin{aligned}
d(e^{\frac{1}{2}\gamma _{i}\phi ^{i}}\ast U^{\gamma
})&=-\frac{1}{2}\gamma _{j}e^{\frac{1}{2}\gamma _{i}\phi
^{i}}d\phi ^{j}\wedge \ast U^{\gamma }\\
\\
&\quad+\sum\limits_{\alpha -\beta =-\gamma }e^{\frac{1}{2} \alpha
_{i}\phi ^{i}}e^{\frac{1}{2}\beta _{i}\phi ^{i}}N_{\alpha ,-\beta
}U^{\alpha }\wedge \ast U^{\beta
},\\
\\
d(\ast d\phi ^{i})&=\frac{1}{2}\sum\limits_{\alpha\in\Delta_{nc}^{+}}^{}\alpha _{i}%
e^{\frac{1}{2}\alpha _{j}\phi ^{j}}U^{\alpha }\wedge e^{%
\frac{1}{2}\alpha _{j}\phi ^{j}}\ast U^{\alpha
}\\
\\
&\quad-e^{\sigma}((H_{i})_{n}^{a}L_{m}^{n}L_{j}^{a})\ast
F^{m}\wedge F^{j},
\end{aligned}
\end{equation}
$\alpha,\beta,\gamma\in\Delta_{nc}^{+}$ and
$[H_{i},E_{\alpha}]=\alpha_{i}E_{\alpha}$ as we have defined
above. The matrices $\{(H_{i})_{n}^{a}\}$ are the ones
corresponding to the generators $\{H_{i}\}$ in the
($N+3$)-dimensional representation chosen.

The field equations corresponding to the scalar Lagrangians of the
symmetric space sigma models which are in the form of \eqref{29.5}
for the dilatons $\{\phi^{i}\}$ and the axions $\{\chi^{m}\}$ are
derived for generic split and non-split scalar cosets $G/K$ in
\cite{ker1,nej1,nej2,ker2}. By referring to \cite{nej2} the
locally integrated first-order field equations for the scalar
Lagrangian \eqref{29.5} can be given as
\begin{equation}\label{219}
\ast \overset{\rightharpoonup }{\mathbf{\Psi }}=-e^{\mathbf{\Gamma }%
}e^{\mathbf{\Lambda }}\overset{\rightharpoonup }{\mathbf{S}}.
\end{equation}
When one applies the exterior derivative on both sides of
\eqref{219} one finds the second-order field equations of the
scalar Lagrangian \eqref{29.5} \cite{nej2}. The column vector
$\overset{\rightharpoonup }{\mathbf{\Psi }}$ is defined as
\begin{equation}\label{220}
\begin{aligned}
\mathbf{\Psi} ^{i}&=\frac{1}{2}d\phi ^{i}\quad\quad\text{for}
\quad\quad i=1,...,r,\\
\\
\mathbf{\Psi} ^{\alpha +r}&=e^{\frac{1}{2}\alpha _{i}\phi
^{i}}\mathbf{\Omega }_{l}^{\alpha}d\chi ^{l}\quad\quad\text{for}
\quad\quad\alpha=1,...,n,
\end{aligned}
\end{equation}
where we have enumerated the roots in $\Delta_{nc}^{+}$. The
vector $\overset{\rightharpoonup }{\mathbf{S}}$ is
\begin{equation}\label{221}
\begin{aligned}
S^{j}&=\frac{1}{2}d\widetilde{\phi}^{j}\quad\quad\text{for}
\quad\quad j=1,...,r,\\
\\
S^{\alpha+r}&=d\widetilde{\chi}^{\alpha}\quad\quad\text{for}
\quad\quad\alpha=1,...,n,
\end{aligned}
\end{equation}
where $\widetilde{\phi}^{j}$, $\widetilde{\chi}^{\alpha}$ are the
five-form Lagrange multiplier dual fields \cite{nej2}. The
matrices $\mathbf{\Gamma}$ and $\mathbf{\Lambda}$ are introduced
as
\begin{equation}\label{222}
\mathbf{\Gamma }%
_{n}^{k}=\frac{1}{2}\phi ^{i}\,\widetilde{g}_{in}^{k}\quad,\quad
\mathbf{\Lambda }_{n}^{k}=\chi ^{m}\widetilde{f}_{mn}^{k},
\end{equation}
where the structure constants $\{\widetilde{g}_{in}^{k}\}$ and
$\{\widetilde{f}_{mn}^{k}\}$ can be read from
\begin{equation}\label{223}
[E_{\alpha },\widetilde{T}_{m}]=\widetilde{f}_{\alpha
m}^{n}\widetilde{T}
_{n}\quad,\quad[H_{i},\widetilde{T}_{m}]=\widetilde{g}_{im}^{n}
\widetilde{T}_{n}.
\end{equation}
The dual generators $\widetilde{T}_{m}$ are defined in
\cite{nej1,nej2} as $\widetilde{T}_{i}=\widetilde{H}_{i}$ for
$i=1,...,r$ and $\widetilde{T}_{\alpha+r}=\widetilde{E}_{\alpha}$
for $\alpha=1,...,n$. Here $\widetilde{H}_{i}$ and
$\widetilde{E}_{\alpha}$ are the associated dual generators of the
dual fields which together with the scalar generators
$\{H_{i},E_{\alpha}\}$ parameterize the dualized scalar coset
\cite{nej1,nej2}. In \cite{nej2} the structure constants
$\{\widetilde{g}_{in}^{k}\}$ and $\{\widetilde{f}_{mn}^{k}\}$ are
calculated as
\begin{subequations}\label{224}
\begin{gather}
\widetilde{f}_{\alpha m}^{n}=0,\quad\quad m\leq r\quad,\quad
\widetilde{f}_{\alpha ,\alpha +r}^{i}=\frac{1}{4}\alpha _{i},\quad\quad%
i\leq r,\notag\\
\notag\\
\widetilde{f}_{\alpha ,\alpha +r}^{i}=0,\quad\quad i>r\quad,\quad
\widetilde{f}_{\alpha ,\beta +r}^{i}=0,\quad\quad i\leq r,\;%
\alpha \neq \beta, \notag\\
\notag\\
\widetilde{f}_{\alpha ,\beta +r}^{\gamma +r}=N_{\alpha ,-\beta },\quad\quad \alpha -\beta =-\gamma,\;\alpha \neq \beta ,\notag\\
\notag\\
\widetilde{f}_{\alpha ,\beta +r}^{\gamma +r}=0,\quad\quad \alpha
-\beta \neq -\gamma ,\;\alpha \neq \beta,\notag\\
\notag\\
\widetilde{g}_{im}^{n}=0,\quad\quad m\leq r\quad,\quad\widetilde{%
g}_{im}^{n}=0,\quad\quad m>r,\; m\neq n,\notag\\
\notag\\
\widetilde{g}_{i\alpha }^{\alpha }=-\alpha _{i},\quad\quad\alpha
>r.\tag{\ref{224}}
\end{gather}
\end{subequations}
By using the first-order formulation of the scalar coset manifold
of the $D=7$ Abelian Yang-Mills supergravity namely
$SO(N,3)/SO(N)\times SO(3)$ given above we can locally integrate
the bosonic field equations \eqref{213} and \eqref{218}. One can
introduce dual fields which are nothing but the Lagrange
multipliers and by using the fact that locally a closed form is an
exact one one can derive the corresponding local first-order field
equations such that the exterior derivative operator can be
extracted on both sides of the equations \eqref{213} and
\eqref{218}. Thus if we introduce the dual three-form
$\widetilde{B}$, the set of dual four-forms
$\{\widetilde{A}^{j}\}$ and the dual five-forms
$\{\widetilde{\sigma},\widetilde{\phi}^{i},\widetilde{\chi}^{m}\}$
we can locally derive the first-order field equations as
\begin{subequations}\label{225}
\begin{gather}
e^{2\sigma}\ast G=-d\widetilde{B},\notag\\
\notag\\
e^{\sigma}\mathcal{M}_{j}^{i}\ast F^{j}=-d\widetilde{A}^{i}-\sqrt{2}\,d\widetilde{B}\wedge\eta_{j}^{i}A^{j},\notag\\
\notag\\
\ast d\sigma=-d\widetilde{\sigma}-\frac{4}{5} B \wedge d\widetilde{B}-\frac{2}{5}\delta_{ij}A^{i}\wedge d\widetilde{A}^{j},\notag\\
\notag\\
e^{\frac{1}{2}\alpha_{i}\phi^{i}}(\mathbf{\Omega})_{l}^{\alpha}\ast
d\chi^{l}=-(e^{\mathbf{\Gamma}}
e^{\mathbf{\Lambda}})_{j}^{\alpha+r}S^{j},\notag\\
\notag\\
\frac{1}{2}\ast
d\phi^{m}=-(e^{\mathbf{\Gamma}}e^{\mathbf{\Lambda}})_{j}^{m}S^{j}+
\frac{1}{2}(H_{m})_{ji}A^{j}\wedge
d\widetilde{A}^{i}+\frac{1}{2\sqrt{2}}\, \eta_{i}^{k}
(H_{m})_{jk}A^{j}\wedge A^{i} \wedge d
\widetilde{B}.\tag{\ref{225}}
\end{gather}
\end{subequations}
 The exterior differentiation of the equations \eqref{225} gives
the second-order equations \eqref{213} and \eqref{218}. If we take
the exterior derivative of the last equation in \eqref{225} we
have to make use of the identities \eqref{26}, \eqref{27} and
\eqref{28}, also the fact that as $H_{i}^{T}\eta=-\eta H_{i}$ and
since $\eta$ is a diagonal matrix $H_{i}\eta$ must have
anti-symmetric matrix representatives, to obtain the corresponding
second-order equation in \eqref{218}. We have also used the
first-order equations \eqref{219} to derive \eqref{225}. In the
second equation above we have used the Euclidean signature metric
to raise the indices.

\section{Dualisation}

In this section we will dualize the bosonic field content of the
$D=7$ Abelian Yang-Mills supergravity and we will construct the
dualized coset element to realize the bosonic field equations
namely \eqref{213} and \eqref{218} by means of the Cartan form of
the dualized coset element. Our formulation will be in parallel
with the one given in \cite{julia2}. We will determine the
structure constants of the Lie superalgebra which parameterize the
dualized coset element so that the second-order bosonic field
equations \eqref{213} and \eqref{218} can be obtained from the
Cartan-Maurer equation which the Cartan form of the dualized coset
element satisfies. Likewise in \cite{julia2} we will also obtain
the locally integrated first-order field equations \eqref{225} as
a twisted self-duality equation of the Cartan form corresponding
to the dualized coset element.

We start by assigning a generator for each bosonic field in
\eqref{22}. The original generators $\{K,V_{i},Y,H_{j},E_{m}\}$
will couple to the fields $\{\sigma,A^{i},B,\phi^{j},\chi^{m}\}$
respectively. The dual generators will also be introduced as
$\{\widetilde{K},\widetilde{V}_{i},\widetilde{Y},
\widetilde{H}_{j},\widetilde{E}_{m}\}$ which will be coupled to
the dual fields
$\{\widetilde{\sigma},\widetilde{A}^{i},\widetilde{B},
\widetilde{\phi}^{j},\widetilde{\chi}^{m}\}$. We remind the reader
of the fact that these dual fields have already appeared in the
first-order field equations in \eqref{225} as a result of the
local integration of the second-order field equations \eqref{213}
and \eqref{218}, they correspond to the Lagrange multiplier fields
in our dualisation construction which is another manifestation of
the Lagrange multiplier method which leads to the first-order
formulation of the theory \cite{pope}. We require that the Lie
superalgebra to be constructed from the original and the dual
generators has the $\mathbb{Z}_{2}$ grading so that the generators
will be chosen as odd if the corresponding potential is an odd
degree differential form and otherwise even \cite{julia2}. In
particular $\{V_{i},\widetilde{K},\widetilde{Y},
\widetilde{H}_{j},\widetilde{E}_{m}\}$ are odd generators and
$\{K,\widetilde{V}_{i},Y,H_{j},E_{m}\}$ are even. As it will be
clear later, the dualized coset element will be parameterized by a
differential graded algebra which is generated by the differential
forms and the generators we have introduced above. This algebra
covers the Lie superalgebra of the field generators. The odd
(even) generators behave like odd (even) degree differential forms
under this graded differential algebra structure when they commute
with the differential forms. The odd generators obey the
anti-commutation relations while the even ones and the mixed
couples obey the commutation relations.

Now let us consider the coset element
\begin{equation}\label{31}
\nu=e^{\frac{1}{2}\phi^{j}H_{j}}e^{\chi^{m}E_{m}}e^{\sigma
K}e^{A^{i}V_{i}}e^{\frac{1}{2}BY}.
\end{equation}
The Cartan form
\begin{equation}\label{32}
\mathcal{G}=d\nu\nu^{-1},
\end{equation}
which is a Lie superalgebra valued one-form can be calculated as
\begin{subequations}\label{33}
\begin{align}
{\mathcal{G}}&=\frac{1}{2}d\phi
^{i}H_{i}+\overset{\rightharpoonup}{{\mathbf{E}}^{\prime
}}\:{\mathbf{\Omega }}\:\overset{\rightharpoonup}{d\chi }+
d\sigma K\notag\\
\notag\\
&\quad +e^{c\sigma}\overset{\rightharpoonup
}{{\mathbf{V}}}e^{{\mathbf{U}}}e^{{\mathbf{B}}}\overset{\rightharpoonup
}{{\mathbf{dA}}}+OY,\tag{\ref{33}}
\end{align}
\end{subequations}
where the three-form $O$ is defined as
\begin{equation}\label{34}
O=\frac{1}{2}e^{a\sigma}e^{\chi^{m}z_{m}}e^{\frac{1}{2}\phi^{n}v_{n}}(dB
+A^{i}\wedge dA^{j}b_{ij}).
\end{equation}
We have defined the yet unknown structure constants as
\begin{subequations}\label{35}
\begin{gather}
[K,Y]=aY\quad,\quad\{V_{i},V_{j}\}=b_{ij}Y,\notag\\
\notag\\
[H_{i},Y]=v_{i}Y\quad,\quad [E_{m},Y]=z_{m}Y.\tag{\ref{35}}
\end{gather}
\end{subequations}
The matrices ${\mathbf{U}}$ and ${\mathbf{B}}$ are
\begin{equation}\label{36}
({\mathbf{U}})_{v}^{n}=\frac{1}{2}\phi^{i}\theta_{iv}^{n}\quad\quad,\quad\quad
({\mathbf{B}})_{n}^{j}=\chi^{m}\beta_{mn}^{j},
\end{equation}
where we have introduced the unknown structure constants as
\begin{equation}\label{37}
[H_{i},V_{n}]=\theta_{in}^{t}V_{t}\quad,\quad[E_{m},V_{j}]=\beta_{mj}^{l}V_{l}.
\end{equation}
We have also defined the commutators
\begin{equation}\label{38}
[K,V_{i}]=cV_{i}.
\end{equation}
In \eqref{33} the row vector $\overset{\rightharpoonup
}{{\mathbf{V}}}$ is defined as ($V_{i}$) and the column vector
$\overset{\rightharpoonup }{{\mathbf{dA}}}$ is ($dA^{i}$). In the
derivation of \eqref{33} we have effectively used the formulas
\begin{equation}\label{39}
\begin{aligned}
de^{X}e^{-X}&=dX+\frac{1}{2!}[X,dX]+\frac{1}{3!}[X,[X,
dX]]+....,\\
\\
e^{X}Ye^{-X}&=Y+[X,Y]+\frac{1}{2!}[X,[X,Y]]+.....
\end{aligned}
\end{equation}
We will now define the dualized coset element as
\begin{equation}\label{310}
\nu^{\prime}=e^{\frac{1}{2}\phi^{j}H_{j}}e^{\chi^{m}E_{m}}e^{\sigma
K}e^{A^{i}V_{i}}e^{\frac{1}{2}BY}
e^{\frac{1}{2}\widetilde{B}\widetilde{Y}}e^{\widetilde{A}^{i}\widetilde{V}_{i}}e^{\widetilde{\sigma}
\widetilde{K}}e^{\widetilde{\chi}^{m}\widetilde{E}_{m}}e^{\frac{1}{2}\widetilde{\phi}^{j}\widetilde{H}_{j}}.
\end{equation}
If we define the Cartan form
$\mathcal{G}^{\prime}=d\nu^{\prime}\nu^{\prime-1}$ then it
satisfies the Cartan-Maurer equation
\begin{equation}\label{311}
d\mathcal{G}^{\prime}-\mathcal{G}^{\prime}\wedge\mathcal{G}^{\prime}=0.
\end{equation}
As it is clear from the dualisation of the maximal supergravities
in \cite{julia2} one can determine the structure constants of the
Lie superalgebra which generates the dualized coset element
\eqref{310} either by calculating \eqref{311} and then comparing
it with the second-order field equations \eqref{213} and
\eqref{218} or one can directly calculate the Cartan form
$\mathcal{G}^{\prime}$ in terms of the unknown structure constants
and then compare the twisted self-duality equation
\begin{equation}\label{312}
\ast\mathcal{G}^{\prime}=\mathcal{SG}^{\prime},
\end{equation}
with the first-order equations \eqref{225}. Here $\mathcal{S}$ is
a pseudo-involution of the Lie superalgebra of the original and
the dual generators. For the field content of the $D=7$ Abelian
Yang-Mills supergravity it can be defined as \cite{julia2,nej1}
\begin{subequations}\label{313}
\begin{gather}
\mathcal{S}Y=\widetilde{Y}\quad,\quad\mathcal{S}\widetilde{Y}=-Y\quad,\quad
\mathcal{S}E_{m}=\widetilde{E}_{m}\quad,\quad\mathcal{S}\widetilde{E}_{m}=-E_{m},\notag\\
\notag\\
\mathcal{S}K=\widetilde{K}\quad,\quad\mathcal{S}\widetilde{K}=-K\quad,\quad
\mathcal{S}H_{i}=\widetilde{H}_{i}\quad,\quad\mathcal{S}\widetilde{H}_{i}=-H_{i},\notag\\
\notag\\
\mathcal{S}V_{i}=\widetilde{V}_{i}\quad,\quad\mathcal{S}\widetilde{V}_{i}=-V_{i}.\tag{\ref{313}}
\end{gather}
\end{subequations}
In our derivation of the algebra structure we will first use the
fact that the Cartan form $\mathcal{G}^{\prime}$ obeys the twisted
self-duality equation \eqref{312} thus by using the identities
\eqref{39} it can be written as
\begin{subequations}\label{314}
\begin{align}
{\mathcal{G}}^{\prime}&=\mathcal{G}-\frac{1}{2}\ast
d\phi^{i}\widetilde{H}_{i}-e^{\frac{1}{2}\alpha_{i}
\phi^{i}}{\mathbf{\Omega} }_{\beta }^{\alpha }\,\ast d\chi ^{\beta
}\widetilde{E}_{\alpha}\notag\\
\notag\\
&\quad-\ast d\sigma
\widetilde{K}-e^{c\sigma}\overset{\rightharpoonup
}{\widetilde{{\mathbf{V}}}}e^{{\mathbf{U}}}e^{{\mathbf{B}}}\ast\overset{\rightharpoonup
}{{\mathbf{dA}}}-\ast O\widetilde{Y},\tag{\ref{314}}
\end{align}
\end{subequations}
where we have defined the row vector $\overset{\rightharpoonup
}{\widetilde{{\mathbf{V}}}}$ as $(\widetilde{{\mathbf{V}}}_{i})$.
In writing \eqref{314} we have also used the fact that the set of
the original generators $O$ and the set of the dual generators
$\widetilde{D}$ obey the general scheme \cite{julia2,nej3,nej4}
\begin{subequations}\label{315}
\begin{gather}
[O,\widetilde{D}\}\subset\widetilde{D}\quad,\quad [O,O\}\subset O,\notag\\
\tag{\ref{315}}\\
[\widetilde{D},\widetilde{D}\}=0.\notag
\end{gather}
\end{subequations}
If we insert the Cartan form \eqref{314} into the Cartan-Maurer
equation \eqref{311} and then compare the result with the
second-order field equations \eqref{213} and \eqref{218} we can
determine the desired structure constants of the original and the
dual field generators. The calculation yields
\begin{subequations}\label{316}
\begin{gather}
[K,V_{i}]=\frac{1}{2}V_{i}\quad,\quad[K,Y]=Y\quad,\quad[K,\widetilde{Y}]=-\widetilde{Y},\notag\\
\notag\\
[\widetilde{V}_{k},K]=\frac{1}{2}\widetilde{V}_{k}\quad,\quad\{V_{i},V_{j}\}=-\frac{1}{\sqrt{2}}\eta_{ij}Y\quad,\quad
[H_{l},V_{i}]=(H_{l})_{i}^{k}V_{k},\notag\\
\notag\\
[E_{m},V_{i}]=(E_{m})_{i}^{j}V_{j}\quad,\quad[V_{l},\widetilde{V}_{k}]=-\frac{2}{5}\delta_{lk}\widetilde{K}
+\frac{1}{2}\underset{i}{\sum
}(H_{i})_{lk}\widetilde{H}_{i},\notag\\
\notag\\
\{V_{k},\widetilde{Y}\}=2\sqrt{2}\,\eta_{k}^{l}\,\widetilde{V}_{l}\quad,\quad[Y,\widetilde{Y}]=\frac{16}{5}\widetilde{K}
\quad,\quad[H_{i},\widetilde{V}_{k}]=-(H_{i}^{T})_{k}^{m}\widetilde{V}_{m},\notag\\
\notag\\
[E_{\alpha},\widetilde{V}_{k}]=-(E_{\alpha}^{T})_{k}^{m}\widetilde{V}_{m},\notag\\
\notag\\
[H_{j},E_{\alpha }]=\alpha _{j}E_{\alpha }\quad ,\quad [E_{\alpha
},E_{\beta }]=N_{\alpha ,\beta }E_{\alpha+\beta},\notag\\
\notag\\
[H_{j},\widetilde{E}_{\alpha }]=-\alpha _{j}\widetilde{E}_{\alpha }\quad ,\quad [%
E_{\alpha },\widetilde{E}_{\alpha }]=\frac{1}{4}\overset{r}{\underset{j=1}{%
\sum }}\alpha _{j}\widetilde{H}_{j},\notag\\
\notag\\
[E_{\alpha },\widetilde{E}_{\beta }]=N_{\alpha ,-\beta }\widetilde{E}%
_{\gamma },\quad\quad\alpha -\beta =-\gamma,\;\alpha \neq
\beta,\tag{\ref{316}}
\end{gather}
\end{subequations}
where we have also included the commutation relations of the
generators of $\mathbf{s}_{0}$. The matrices ($(H_{m})_{i}^{j}$,
$(E_{\alpha})_{i}^{j}$) are the matrix representatives of the
corresponding generators ($H_{m},E_{\alpha}$). Also the matrices
($(H_{m}^{T})_{i}^{j}$, $(E_{\alpha}^{T})_{i}^{j}$) are the matrix
transpose of ($(H_{m})_{i}^{j}$, $(E_{\alpha})_{i}^{j}$). The
scalar generators and their duals which are coupled to the
five-form dual fields $\widetilde{\phi}^{j}$ and
$\widetilde{\chi}^{m}$ namely the generators
$\{H_{i},E_{m},\widetilde{E}_{m},\widetilde{H}_{i}\}$ constitute a
subalgebra and their algebra is already constructed for a generic
scalar coset $G/K$ in \cite{nej2}. The commutators and the
anti-commutators which are not listed in \eqref{316} vanish. Now
we can explicitly calculate the Cartan form $\mathcal{G}^{\prime}$
since we have obtained the algebra structure which generates the
coset element $\nu^{\prime}$. Thus using the identities
\eqref{39}, the commutators and the anti-commutators of
\eqref{316} the calculation of the dualized Cartan form yields
\begin{equation}\label{317}
\begin{aligned}
\mathcal{G}^{\prime}&=d\nu^{\prime}\nu^{\prime-1}\\
\\
&=\frac{1}{2}d\phi^{i}H_{i}+e^{\frac{1}{2}\alpha_{i}\phi^{i}}U^{\alpha}E_{\alpha}
+d\sigma K+e^{\frac{1}{2}\sigma}L_{i}^{k}dA^{i}V_{k}+\frac{1}{2}e^{\sigma}GY\\
\\
&\quad+\frac{1}{2}e^{-\sigma}d\widetilde{B}\widetilde{Y}+(\frac{4}{5}B\wedge
d\widetilde{B}+ \frac{2}{5}A^{j}\wedge
d\widetilde{A}^{i}\delta_{ij}+d\widetilde{\sigma})\widetilde{K}\\
\\
&\quad+\overset{r}{\underset{m=1}{\sum}}((e^{\mathbf{\Gamma}}e^{\mathbf{\Lambda}})_{j}^{m}
S^{j}-\frac{1}{2}(H_{m})_{ji}A^{j}\wedge
d\widetilde{A}^{i}\\
\\
&\quad-\frac{1}{2\sqrt{2}}\, \eta_{i}^{k} (H_{m})_{jk}A^{j}\wedge
A^{i} \wedge
d\widetilde{B})\widetilde{H}_{m}+\underset{\alpha\in\Delta_{nc}^{+}}{\sum}(e^{\mathbf{\Gamma}}e^{\mathbf{\Lambda}})_{j}^{\alpha+r}
S^{j}\widetilde{E}_{\alpha}\\
\\
&\quad+(e^{-\frac{1}{2}\sigma}((L^{T})^{-1})_{k}^{l}d\widetilde{A}^{k}+\sqrt{2}\;e^{-\frac{1}{2}\sigma}((L^{T})^{-1})_{k}^{l}
\,\eta_{i}^{k}\,A^{i}\wedge d\widetilde{B})\widetilde{V}_{l}.
\end{aligned}
\end{equation}
If we apply the twisted self-duality equation \eqref{312} on
\eqref{317} we can find the locally integrated first-order field
equations as
\begin{subequations}\label{318}
\begin{gather}
\frac{1}{2}e^{\sigma}\ast G=-\frac{1}{2}e^{-\sigma}d\widetilde{B},\notag\\
\notag\\
e^{\frac{1}{2}\sigma}\L_{j}^{i}\ast
dA^{j}=-e^{-\frac{1}{2}\sigma}((L^{T})^{-1})_{j}^{i}d\widetilde{A}^{j}-\sqrt{2}\;e^{-\frac{1}{2}\sigma}
((L^{T})^{-1})_{j}^{i}\,\eta_{k}^{j}\, A^{k}\wedge d\widetilde{B},\notag\\
\notag\\
\ast d\sigma=-d\widetilde{\sigma}-\frac{4}{5}B\wedge d\widetilde{B}-\frac{2}{5}\delta_{ij}A^{j}\wedge d\widetilde{A}^{i},\notag\\
\notag\\
e^{\frac{1}{2}\alpha_{i}\phi^{i}}(\mathbf{\Omega})_{l}^{\alpha}\ast
d\chi^{l}=-(e^{\mathbf{\Gamma}}
e^{\mathbf{\Lambda}})_{j}^{\alpha+r}S^{j},\notag\\
\notag\\
\frac{1}{2}\ast
d\phi^{m}=-(e^{\mathbf{\Gamma}}e^{\mathbf{\Lambda}})_{j}^{m}S^{j}+
\frac{1}{2}(H_{m})_{ji}A^{j}\wedge
d\widetilde{A}^{i}+\frac{1}{2\sqrt{2}}\, \eta_{i}^{k}
(H_{m})_{jk}A^{j}\wedge A^{i} \wedge d
\widetilde{B}.\tag{\ref{318}}
\end{gather}
\end{subequations}
These equations are the same equations with the first-order
equations \eqref{225}. Thus as we have intended we have obtained
the locally integrated first-order equations \eqref{225} through
the twisted self-duality equation \eqref{312} which the Cartan
form $\mathcal{G}^{\prime}=d\nu^{\prime}\nu^{\prime-1}$ satisfies.
This result also verifies the validity of the algebra structure
derived in \eqref{316} which generates the dualized coset element
\eqref{310} that yields the second-order field equations
\eqref{213} and \eqref{218} in the Cartan-Maurer equation
\eqref{311}.
\section{Conclusion}
We have performed the bosonic dualisation of the $D=7$ Abelian
Yang-Mills supergravity \cite{koh} which is the low energy
effective limit of the fully Higgsed $D=7$ heterotic string. We
have discussed the coset construction of the scalars which belong
to the coupling vector multiplets in section two. After deriving
both the second and the first-order field equations in section two
we have generalized the coset structure of the scalars to the
entire bosonic sector by dualizing the original bosonic fields and
then by determining the Lie superalgebra which generates the
dualized coset element in section three. Therefore we have
reformulated the bosonic sector of the $D=7$ matter coupled
supergravity as a non-linear sigma model. We have also obtained
the first-order formulation of the bosonic sector of the $D=7$
Abelian Yang-Mills supergravity in the form of a twisted
self-duality equation.

The dualisation method we have used in this work which is another
manifestation of the Lagrange multiplier methods \cite{pope} is an
extension of the one introduced in \cite{julia2} to the matter
coupled supergravities. Since the scalar coset manifold
$SO(N,3)/SO(N)\times SO(3)$ we are dealing with in this work is in
general a non-split type depending on the number of coupling
multiplets we have used the results of \cite{nej2} which presents
the dualisation of a generic non-split scalar coset. Thus this
work also serves as an application of the achievements of
\cite{nej2}.

Although we have derived the algebra which can be considered as a
gauge to generate the dualized coset element, so that it realizes
the bosonic field equations within the context of a non-linear
sigma model, the group theoretical construction of the coset is to
be studied. Furthermore one can also construct the dualized
Lagrangian and then determine the global and the local symmetry
groups.

The symmetries of the supergravity theories have been studied in
recent years to gain insight in the symmetries and the duality
transformations of the string theories. Especially the global
symmetries of the supergravities contribute to the knowledge of
the non-perturbative U-duality symmetries of the string theories
and the M theory \cite{nej125,nej126}. An appropriate restriction
of the global symmetry group $G$ of the effective low energy limit
supergravity theory to the integers ${\Bbb{Z}}$, namely
$G({\Bbb{Z}})$, is conjectured to be the U-duality symmetry of the
relative string theory \cite{nej125}. The Lie superalgebra we have
constructed in section three which generates the dualized coset
element is a parametrization of the coset structure $G/K$ of the
bosonic sector thus it contains the necessary information about
the enlarged symmetry group $G$ of the $D=7$ matter coupled
supergravity. Therefore as we have mentioned above the improved
global symmetry analysis of the $D=7$ matter coupled supergravity
may also reveal the symmetry scheme of the $D=7$ heterotic string.

As we have discussed in section two the $T^{3}$-compactification
of the $D=10$ type I supergravity that is coupled to the
Yang-Mills theory \cite{d=10,tani15} gives the $D=7$ matter
coupled supergravity, also an equivalent dual bosonic Lagrangian
of the $D=7$ Abelian Yang-Mills supergravity in which the two-form
potential $B$ is replaced by a dual three-form field
\cite{town1,town2} corresponds to the $K_{3}$ compactification
\cite{townref25} of the $D=11$ supergravity which is conjectured
to be the low energy limit of the M theory. Thus by constructing a
tool to study the general symmetries of the $D=7$ Abelian
Yang-Mills supergravity we have also contributed to the symmetry
studies of these higher dimensional theories especially the M
theory. In \cite{town3} the string-membrane dualities in $D=7$
which arise from the comparison of the construction of the $D=7$
Abelian Yang-Mills supergravity either as a toroidally
compactified heterotic string or a $K_{3}$-compactified $D=11$
supermembrane are discussed. It is possible that the complete
dualisation and the non-linear realization of the bosonic sector
of the $D=7$ Abelian Yang-Mills supergravity will also help to
understand the string-membrane and the string-string dualities in
$D=7$.

One can also extend the construction of this work by including the
gravity sector as well, in a way presented in
\cite{d=83,d=84,d=85} and then seek for the interpretation of the
gravity-included dualized coset as a non-linear realization of a
Kac-Moody global symmetry group.

\end{document}